\documentclass[12pt]{JHEP}

\usepackage{amsmath,epsfig}
\usepackage{amssymb,amsfonts}
\usepackage{latexsym}
\usepackage{longtable}

\relax
\renewcommand{\theequation}{\arabic{section}.\arabic{equation}}
\def\be{\begin{equation}}
\def\ee{\end{equation}}
\def\bea{\begin{eqnarray}}
\def\eea{\end{eqnarray}}

\newcommand\fverb{\setbox\pippobox=\hbox\bgroup\verb}
\newcommand\fverbdo{\egroup\medskip\noindent%
                        \fbox{\unhbox\pippobox}\ }
\newcommand\fverbit{\egroup\item[\fbox{\unhbox\pippobox}]}

\newcommand{\bear}{\begin{eqnarray}}

\newcommand{\eear}{\end{eqnarray}}

\newbox\pippobox

\def\N{{\cal N}}

\def\6{\partial}

\def\a{\alpha}

\def\half{\frac12}

\def\sq
\def\a{\alpha}

\def\hri#1#2{\href{http://arxiv.org/abs/#1}{[ArXiv:#1]#2}}
\def\hre#1#2{\href{http://arxiv.org/abs/#1/#2}{[ArXiv:#1/#2]}}

\def\Zbf{{\bf Z}}

\title{Free Fermion Orientifolds}
\author{Elias  Kiritsis$^{1}$, Michael Lennek$^{2}$ and Bert  Schellekens$^{3,4,5}$\\~\\~\\
$^1$Department of Physics, University of Crete,
71003 Heraklion, Greece\\~\\
$^2$CPHT, Ecole Polytechnique,
 91128, Palaiseau, France,
 ( UMR du CNRS 7644).\\
~\\
$3$NIKHEF,
Kruislaan 409, 1009DB Amsterdam,
The Netherlands\\
~\\
$^4$IMAPP, Radboud Universiteit Nijmegen, The Netherlands
~\\
$^5$Instituto de F\'\i sica Fundamental, CSIC, Madrid, Spain}

\preprint{CPHT-RR081.1008\\
NIKHEF/2008-029}

\abstract{We investigate a class of orientifold models based on tensor products
of 18 Ising models. Using the same search criteria as for the comparable case of
Gepner model orientifolds we find that there are no three-family standard model
configurations with tadpole cancellation. Even if we do not impose the latter requirement,
we only find one such configuration in the special case of complex free fermions. In order
to allow a comparison with other approaches we enumerate the Hodge numbers of the
type-IIB theories we obtain. We provide indications that there are fermionic IIB vacua that are not $Z_2\times Z_2$ orbifolds.}

\begin{document}

\section{Introduction}
\label{intro}

In the last decade, there has been substantial progress in the construction of semi-realistic, standard-model-like
 string spectra using orientifolds.  It was realized early on that orientifolds are successfully tuned to allow
  bottom up constructions of the SM spectrum using D-branes, \cite{akt,ibanez}. This has led to a separation of
  the problem of the construction of SM-like vacua to that of a local problem (engineering the SM stack of branes)
   and a global problem (tadpole cancellation).

Two classes of approaches have been applied to the construction of orientifold vacua, namely geometric and algebraic.
  The former starts with torus compactifications, to which orbifold and orientifold projections are applied.
   The latter starts with some rational conformal field theory (RCFT) to which boundary and crosscap states are added.
    In general, geometric constructions have the advantage that the moduli space of a solution
    is under much better control, whereas the algebraic approach probes deeper into the landscape of
     possibilities. The geometric approach has so far been applied mainly to $\Zbf_2 \times \Zbf_2$,
      $\Zbf_6$
      and $\Zbf_6'$ orientifolds (see  \cite{Blumenhagen:2005mu,Gmeiner:2008xq} and references therein).

The algebraic approach has been applied successfully to Gepner Models \cite{Gepner:1987qi}. It gave the richest class of SM-like
 vacua without chiral exotics \cite{dhs}. Moreover it also gave the richest class of possibilities of embedding
  the SM spectrum into Chan-Paton groups, \cite{adks}. For other work on Gepner orientifolds see
  \cite{Angelantonj:1996mw}-\cite{Aldazabal:2004by}.

In principle, the geometric and algebraic (RCFT) approaches are not strictly separated. Here we will consider
 a class of orientifold vacua that is accessible from both of these directions, namely orientifolds of free-fermionic
  theories. From the algebraic point of view, this class is obtained by tensoring 18 Ising models in order to obtain
   the required central charge of 9, and imposing a world-sheet supersymmetry constraint. Geometrically, it is known
    that such theories are closely related to $\Zbf_2 \times \Zbf_2$ orientifolds. Our hope is, on the one hand,
    to find a standard-model-like configuration that can be studied from both perspectives.  On the other hand,
     such a configuration might allow an explicit computation of couplings in a realistic example. While this
      is in principle possible for tensor products of $\N=2$ minimal models ({\it i.e.} Gepner models), the
       required formalism is in practice only available for the simplest RCFT, the Ising model, or for the free boson.

The fermionic construction of string compactifications was pioneered in the heterotic context, \cite{abk1}-\cite{abk3}.
 It has proved a very practical tool and the phenomenologically most successful heterotic
vacua were found in this context, \cite{nahe}-\cite{far}. It allowed for an algorithmic search of vacua using computers,
 and a rather straightforward algorithmic computation of the superpotential that has been exploited up to eighth
  order in the fields \cite{atr}. The fermionic approach to the heterotic string has been revived recently,
  \cite{fk1}-\cite{fk3}.  It has been also used for statistical studies of the heterotic landscape, \cite{dienes1}-\cite{dienes4}.

  The art of free-fermion model building consists of simultaneously satisfying three requirements:
  world-sheet supersymmetry, modular invariance and, if desired, space-time supersymmetry. The first
  and the latter condition are essentially always satisfied in the same way. World-sheet supersymmetry
  is imposed by using a realization of the world-sheet  supercurrent first presented in \cite{abk1}, leading
  to a ``triplet constraint" on the free fermions, which in the language of conformal field theory results
  in extending the chiral algebra by certain currents of spin 3. Space-time supersymmetry always
  amounts to an extension of the chiral algebra by a definite spin-1 current. However, there are various
  ways of dealing with the third constraint, modular invariance. The most general one, proposed in
  \cite{klt1} and  \cite{klt2} is to derive conditions on the boundary conditions of fermions on non-contractible
  cycles on the torus and higher genus surfaces (dealing with higher loop modular invariance
  is not entirely straightforward, however \cite{Kawai:1987ew}).  The second one is to consider the special
  situation where free fermion and free boson constructions overlap, {\it i.e.} complex free fermion pairs, in which
  case one may use the covariant lattice construction \cite{LLS}, and modular invariance at arbitrary genus
  can be derived using Lorentzian self-dual lattices. The third method is to use simple current
  modifications of diagonal partition functions, in which case consistency is
  guaranteed by general theorems \cite{Fuchs:2004dz}. The choice of method is limited by the
  requirement of being able to perform an orientifold projection on the result. For the first method this
  problem was studied in \cite{Bianchi:1989du,Bianchi:1990yu,Bianchi:1990tb}, but so far no fully
  general method has been formulated.  For the other two methods such a method {\it does} exist. As we shall see in the
  next section, the simple current method in combinations with the requirement of space-time
  supersymmetry does require bosonization of some, but not all of the fermions. It is thus somewhat less general
  than the full free fermion construction, but more general than a free boson construction, and it
  is, to the best of our knowledge, the most
  general method currently available for free fermion orientifold constructions.

Although the  $\Zbf_2\times \Zbf_2$ orbifolds relevant for the fermionic constructions have been successful in
the heterotic context the associated  results for $\Zbf_2 \times \Zbf_2$ orientifolds have not been very
 encouraging so far, \cite{billion,mrd}.  It should be however be appreciated that   the searches done so
  far concern a rather small set of possible standard model realizations. In \cite{adks} a minimally biased
   set of requirements was formulated, which allows many more -- although sometimes rather exotic -- realizations
    of the standard model. Basically, the only requirement (constraint) is that all quarks and leptons originate
     from a maximum of four participating branes, and that the strong and weak gauge groups are not diagonally
      embedded in multiple brane stacks. This allows for example arbitrary embeddings of the weak hypercharge
       $Y$ and quarks and leptons originating from rank two tensors (that were classified), as well as various kinds of gauge unification. Here we will use exactly
          the same set of requirements.
For a more detailed description we refer to  \cite{adks}.

Our main conclusion regarding standard model spectra is that even with these much broader search criteria,
 the set of free fermion orientifolds (and hence presumably the $\Zbf_2 \times \Zbf_2$ orientifolds) is
  an extremely poor region in the orientifold landscape, in comparison to orientifolds of interacting CFT's,
   in particular of Gepner models. Since we use identical search criteria in both cases this is a fair comparison.
     Indeed, in the present search we did not find any solution to the tadpole conditions that contains the
      standard model spectrum. Even keeping only the condition that the spectrum is right, before trying to
       find a tadpole canceling hidden sector, we found just a handful of solutions of two different chiral
       types (which, however, are remarkably simple and elegant). By contrast, in the case of Gepner models
       both problems (finding the standard model spectrum with or without tadpole cancellation) had a huge number
       of solutions:  the number of distinct chiral types in that search was more than 19000 \cite{adks},
       in comparison with just two in the present search.

This paper is organized as follows. In the next section we will describe the free fermion CFT's we are considering.
 In section three we discuss the closed sector of these CFT's, and present a list of Hodge numbers for comparison
 with other work. This list should in particular be useful to determine the precise scope of our search. Since we
 do not have any formalism to deal with free fermion orientifolds in full generality ({\it i.e.} for 18 unpaired real fermions),
 it would be interesting to know the full list of Hodge data for the general case, and compare with ours. Despite the
 more than twenty years of history of the subject, apparently such a list is not available at present.
 Finally, in section four we will present the standard model search results. The appendix contains a more detailed
 list of Hodge data, including results on heterotic singlets and the number of boundary states.

\section{CFT considerations}
\label{CFT}
Our basic building block is the Ising CFT, which has three primaries $0$, $\psi$ and $\sigma$ with
conformal weights $0$, $\half$ and $\frac{1}{16}$ respectively. Since its central charge is $\half$ one
can tensor 18 copies in order to obtain a $c=9$ ``internal"  CFT for a compactified type-II string theory.
Fermionic string theory consistency requires an $\N=1$ world-sheet supersymmetry. Unlike the building blocks
 used in Gepner models, the $\N=2$ minimal models, the Ising building blocks are not supersymmetric. But it
  has been known for a long time \cite{abk1,klt1,abk2} how to realize world-sheet supersymmetry
   on a triplet of Ising models. The world sheet supercurrent is simply the product of the three fermionic
   currents of the factors, $\psi_1\psi_2\psi_3$.  Having realized supersymmetry on  a triplet of fermions,
   we still have to impose it on products of supersymmetric building blocks, so that their NS and R sectors
   are properly aligned. This is done, as in the case of Gepner models, by extending the chiral algebra with
   all products of the supercurrents of the building blocks, including the space-time NSR factor.  These products
    are sometimes  called ``alignment currents". They have spin 3, because they are products of
    spin-$\frac32$ of the separate factors, or the supercurrent $X^{\mu}\partial \psi_{\mu}$ of the
    NSR factor of the theory.
       Extending the algebra by these currents
    implies a projection on the spectrum, which in the special case of free fermionic models is called the ``triplet constraint".

In the case of interest, one divides the 18 Ising models into six groups of three to impose this constraint.
The result is a  fermionic string theory, which in general has a spectrum without space-time supersymmetry.
 To obtain space-time supersymmetry we have to perform another extension of the chiral algebra, by a spin-1
 current that is spinorial in the  NSR sector. The resulting projection on the spectrum is of course the GSO-projection.
 This current consists of an NSR spin fields with weight $\frac{5}{8}$ combined with six Ising spin  fields $\sigma$,
 so that the total conformal weight is 1. Locality with the  alignment currents requires that  there be an odd number
 of $\sigma$ fields in each fermionic triplet, and then obviously the only solution is to choose precisely one per triplet.

There is an important difference between the alignment currents and the space-time supercurrent. The
former consists entirely of simple currents, whereas the latter involves the Ising field $\sigma$, which
is not a simple current. The boundary state formalism we want to use is the one of \cite{Fuchs:2000cm}, which
includes the most general available extension of earlier work of the Rome group
\cite{Pradisi:1996yd,Pradisi:1995pp}, which in its turn is based on the classic paper by Cardy \cite{Cardy:1989ir}.
This formalism produces the complete set of boundary states for all simple current extensions of the
chiral algebra. Unfortunately it cannot be applied to extensions that are not simple current related, like the space-time supercurrent we
encounter here.

But there is a way out of this in some cases. An Ising model corresponds to a real (Majorana) free fermion.
 If we combine a pair of them into a complex free fermion, then the spinor current turns out to be a  simple
  current. Such a pairing implies that the two fermions have the same boundary conditions on any cycle on any
  Riemann surface, and hence is a restriction on the total number of possibilities. This can be achieved by
  extending the chiral algebra of the theory with the spin-1 current $\psi_i \psi_j$, where $i$ and $j$ label
   the fermions to be paired. In order to use the simple current boundary state formalism we have to group the
    six fermions participating in the space-time supercurrent into  three pairs.
This yields then a  type-II theory built out of three  complex fermions (with standard, periodic and anti-periodic
 boundary conditions) and twelve real fermions.  We may consider pairing some of the remaining real fermions as well.
 Such a pairing replaces the real fermion pair by a free boson compactified on a circle of radius $R^2=4$. The resulting CFT has central charge 1 and may be thought of as
the extrapolation of the $D_n$ affine Lie algebras to $n=1$.  Therefore we will denote it as $D_1$. Hence the resulting $c=9$ CFT is
in  general built out of a combination of Ising models and free bosons. This should not be confused with the case studied in
 \cite{Kiritsis:2008ry}, the $2^6$ Gepner model. This models are also tensor products of free fermions and free
 bosons, but in this case the bosons are on a circle of radius $R^2=8$, and are not straightforwardly related to free fermions.

It may seem that there is no advantage to pairing two real fermions into a boson.  Normally, the pairing 
of fermions reduces the number of options for choosing fermion boundary conditions, and hence the 
largest number of free-fermionic CFT's is obtained by leaving all boundary conditions free and independent. 
Indeed, the pairing of two fermions amounts to an extension of the chiral algebra. In general, there are two 
ways of dealing with such extensions. The first is to extend the chiral algebra directly, and work with the
 reduced set of characters this implies. The second is to implement the extension as a modular invariant
  partition function (MIPF), which has the form of a sum squares of linear combinations of characters.  
  These linear combinations correspond to the reduced set of characters  of the extended chiral algebra, 
  and indeed this MIPF is identical to the diagonal partition function of the extended theory. These two 
  methods therefore yield identical closed string sectors. We will refer to these two cases as a {\it direct 
  extension} and a {\it MIPF extension} henceforth. Although they yield identical closed strings, there is an
   important different between these two cases when open strings are considered, using the formalism of
    \cite{Fuchs:2000cm}. In the case of a direct extension, only boundary states are allowed that respect 
    the extended symmetry, whereas in the case of a MIPF extension only the original chiral algebra is required 
    to be respected. Hence in that case there are boundary states that respect the extension, but also additional
     ones that do not respect it. Therefore it is in general advantageous to implement an extension as a MIPF,
      unless the extended symmetries are themselves required for the physics of the problem under consideration. 
       The latter is true for world-sheet supersymmetry, sometimes for space-time supersymmetry, but not for the
        pairing extension discussed above.

However, there is one exception to the foregoing if the formalism of \cite{Fuchs:2000cm} is used. This
exception occurs when the extended CFT has simple currents that result from fixed point resolution.
In that case working directly in the extended CFT allows us to use these simple currents to build
MIPFs that cannot be obtained as simple current MIPFs in the unextended theory. In the unextended
theory those MIPFs are exceptional invariants, to which the general formalism of \cite{Fuchs:2000cm}
does not apply, and for which ad-hoc formalisms must be developed, as was done for example
for the E-type invariants of $SU(2)$ \cite{Behrend:1999bn}.

This can most easily be studied in the tensor product of two fermions. This has a total of nine primaries,
four of which are simple currents. If we extend the chiral algebra by the spin-1 current $\psi_1 \psi_2$, we
get a new CFT with four primaries. Two of these  are the identity $(0,0) + (\psi, \psi)$ and the free fermion
$(0,\psi) + (\psi, 0)$. The other two originate from the combination $(\sigma,\sigma)$. This turns out to be
a fixed point of the extension current  $\psi_1 \psi_2$, which is resolved into two separate fields in the
extended CFT.  In rare cases it may happen that such a resolved fixed point field becomes a simple
current in the extended CFT, and this is such a case.
If we consider the MIPF obtained by using the simple current $\psi_1 \psi_2$, we get a total of six boundary
states. Four of these respect the  extended symmetry, and two of them do not. If we work instead directly in
the extended CFT, {\it i.e.} $D_1$, we only see the four boundary states the respect the extension.
So here the MIPF has the advantage over the direct extension. To see the opposite, consider 16
free fermions. We can pair these, using a direct extension, into 8 free bosons. This CFT, $(D_1)^8$, has
a simple current MIPF corresponding to $D_8$, which is also a simple current MIPF of $({\rm Ising})^{16}$, but it also
has a MIPF corresponding to $E_8$, which is {\it not} a simple current MIPF of $({\rm Ising})^{16}$. Although we
would not be able to obtain this $E_8$ theory with simple current MIPFs of only Ising models, it can be obtained
with the method we use in the present paper, namely combinations of Ising models and free bosons. In fact, although
we would expect that examples exist which can only be obtained using the full free fermion construction, and not
by means of simple currents in combinations of free boson and free fermion CFT's, we are not aware of any such
example.

The conclusion is that to maximize the number of cases we are able to consider with the formalism at
our disposal, we should consider all possible options for pairings of the 12 remaining free fermions. The starting
point is the completely unpaired case. This is a CFT that is a tensor product of a four-dimensional
NSR model, three free bosons $\phi$ labeled $a,b,c$, and twelve free fermions $\psi$ labeled $1,\ldots,12$. The
chiral algebra is extended by the following alignment currents
\be
\label{align}
\partial X_{\mu} \psi^{\mu}  e^{i\phi_a} \psi_1 \psi_2 \ \ \ \
\partial X_{\mu} \psi^{\mu}  e^{-i\phi_a} \psi_3 \psi_4
\ee
plus four more with labels $(b,5,6), (b,7,8), (c,9,10), (c,11,12)$. Here $\psi^{\mu}$ are the NSR fermions.
The space-time supersymmetry current
is $S_{\alpha} \sigma_a \sigma_b \sigma_c$, where $S_{\alpha}$ denote the NSR spin fields combined
with the usual contribution from the bosonized superghosts, and $\sigma_a, \sigma_b$ and $\sigma_c$
are the $D_1$ spinors. The latter three are simple currents, and for all practical purposes, so is the
NSR spin field. The nicest way of dealing with it explicitly as a simple current is to use the covariant lattice method
of \cite{LLS}, where it becomes a spinor of $D_5$. Note that there are two choices available for each
of the factors of the space-time supersymmetry current, but all these choices are equivalent.

All other options are obtained from this starting point by adding pairing currents $\psi_i \psi_j$,
with $i,j = 1, \ldots 12$. These pairing currents are always local with respect to the alignment currents, the
space-time susy current and with respect each other, so they can be added without any constraint. However,
we have to close the algebra after adding any such current, which may lead to undesirable consequences.

Let us first consider the special case where we only add pairing currents for the first four fermions. If we add just
one pairing current, the distinct possibilities (taking permutations into account )
are $\psi_1 \psi_2$ and $\psi_1\psi_3$. The former choice, when
combined with (\ref{align}), implies an extension of the chiral algebra with
$\partial X_{\mu} \psi^{\mu}  e^{i\phi_a}$, which means that the four-dimensional NSR model
is extended to a six-dimensional one. Hence all theories we get this way are torus compactifications
of a six-dimensional theory. This is of no interest, since in such a theory all characters are non-chiral
in space-time
and hence there is no possibility for obtaining the standard model from boundary states\rlap.\footnote{Note that
we are considering direct extensions here.  If, on the other hand, we implement the extension by $\psi_1 \psi_2$ as a MIPF, there is a possibility of having a
six-dimensional bulk CFT but space-time chiral boundary states that do not respect the bulk symmetry.} If we
add the current  $\psi_1\psi_3$, then closure of the algebra with the two currents in  (\ref{align}) implies
that also the combination $\psi_2\psi_4$ is in the chiral algebra.
Hence there are just two options that are
of interest, namely no pairing, and the pairing $(1,3)(2,4)$.

One may continue this procedure to eight fermions. Obvious solutions are no pairings, two pairings
(either (1,3)(2,4) or (5,7)(6,8), which are equivalent under permutation), or four pairings, (1,3)(2,4)(5,7)(6,8).
But in addition to these three distinct possibilities one may also consider pairings between the first
and the second group of four. Here one also encounters some possibilities that are of no interest. For example
the pairing $(1,5)(1,6)$ has the effect of combining the three fermions (1,5,6) into $SO(3)$. But this has
no advantages, because $SO(3)$ has no simple currents on top of those of the original free fermions.
Similarly, extensions to other $SO(N)$ groups with $N\geq 3$ need not be considered.
Taking this, as well as all permutations, into account, we arrive at a total of 11 possibilities, including the
three described above.

We proceed in a similar way with the case of twelve fermions. Here we obtain a total of 62 distinct cases,
including all combinations of four and eight fermion sub-cases. There is some overcounting in this set, because
it turns out that some purely free boson (complex fermion)
cases are extensions of others by currents of spin two or three, which
is clearly a direct extension that has no advantages over a MIPF extension.
There may be other such ``useless" extensions for mixtures
of real and complex fermions, but there is an important
caveat here: there are examples that look like extensions of cases that are not on the list of 62
themselves.
 We found that if this happens the ``de-extended" combination does not have a valid world-sheet
supersymmetry realization, even though the extended combination does. This can happen because
the aforementioned
spin two or three current may be combinations of pairing currents and world-sheet supersymmetry
currents, and removing it may therefore destroy world-sheet supersymmetry.  Since the
superfluous cases are anyway among the easiest to deal with in terms of computer time,
it was not worthwhile to eliminate them.

\section{The closed string sector and the associated geometry}
\label{CY}
As discussed earlier, we are utilizing a purely algebraic approach in order to construct these models.
 The fact that there are no explicit geometric considerations that enter into the model construction
  method makes examining the resulting compactification geometries interesting.  The primary constraints
   on what compactification geometries result stem only from the CFT considerations discussed in Sect.~\ref{CFT}
    and stringy consistency conditions.  We shall start our discussion of the compactification geometries at
     the ``global'' level by discussing all of the compactification geometries found.

In the course of this study, we found thirty-two different compactification manifolds.  These manifolds
 are differentiated solely on the basis of their Hodge numbers (namely $h_{11}$ and $h_{12}$) and the amount
  of space-time supersymmetry preserved.  The full list of manifolds is presented on Table \ref{Hodge}.
    As the structure of the table suggests, we find that each model has a mirror, but we shall defer a
     discussion of mirror symmetry until later.  The table does illustrate that we find a wide variety
     of different Hodge numbers and find that within this set of manifolds there is a large variation
      in the amount of supersymmetry preserved.

\begin{table}
\begin{center}
\begin{tabular}{||c|c|c|c||}
\hline
\hline
Hodge Numbers & \multicolumn{3}{c||}{Amount of $d=4$ SUSY}\\
\cline{2-4}
$(h_{11}, h_{12})$ &$\N=1$  & $\N=2$ & $\N=4$\\
\hline
(51,3) and (3,51) & X & &  \\
(31,7) and (7,31) & X & & \\
(27,3) and (3,27) & X & &  \\
(25,1) and (1,25) & X & &  \\
(21,9) and (9,21) &X & &  \\
(19,7) and (7,19) & X & &  \\
(17,5) and (5,17) & X & &  \\
(15,3) and (3,15) & X & &  \\
(12,6) and (6,12) & X & &  \\
\hline
(21,21)  &  & X &  \\
(19,19) & X & &  \\
(15,15) & X & &  \\
(13,13)& X &X &  \\
(11,11) & X & &  \\
(9,9) & X &X &X  \\
(7,7) & X & &  \\
(5,5) & X &X &  \\
(3,3) & X & &  \\
 (1,1) &  &X &  \\
\hline
\hline
\end{tabular}
\end{center}
\caption{The compactification manifolds found in this study along with the amount of space-time supersymmetry that they preserve.}
\label{Hodge}
\end{table}

As discussed in Sect.~\ref{CFT}, our study consisted of sixty-two different model classes.
 The manifolds listed on Table~\ref{Hodge}, were distributed amongst these different model classes.
   We shall now examine how these manifolds were distributed amongst the different model classes.
    This can give some idea how generically these manifolds may be found in this context.
    There is a large variation between different compactification manifolds
      with respect to the number of model classes that realize them.  The manifold preserving $\N = 4$
       supersymmetry, which most likely corresponds to a toroidal compactification, is found in every
        single model class.  There were also two manifolds preserving $\N =2$ supersymmetry (namely
        $(h_{11},h_{12})=(13,13), (5,5)$) which were each found in over fifty of the model classes.
        There are many model classes which {\it only} realize these three common manifolds.  On the other
         extreme, there are three manifolds, ($\N=1~(25,1), (1,25),(13,13)$) that are only realized in one
          model class each.  These manifolds only are found in the case of the extension only involving powers
           of $D_1$ (that is, all fermions paired).  Most of the manifolds are realized in a relatively small
            number of model classes with twenty-four of the manifolds only being realized in less than a third
            of the available model classes.

Another quantity that can be utilized to differentiate between compactification manifolds is the number of so-called
 ``heterotic singlets''.  The model construction method utilized allows for the counting of the number of massless
  states which transform as singlets under an $E_6$ factor within the chiral algebra\footnote{For toroidal
  compactifications, the $E_6$ factor is enhanced to an $E_8$.  Thus, for these compactifications we count the number
   of $E_8$ singlets instead of $E_6$ singlets.  For this study, this only affects the manifolds preserving $\N=4$ supersymmetry.}.
   The name ``heterotic singlets" derives from the following fact: any type-II partition function with (1,1) space-time supersymmetry can be uniquely mapped
   to a heterotic vacuum with N=1 space-time susy with $E_6$ symmetry via a modular invariance preserving map
   first described in \cite{ENS,LLS} and applied to map type-II strings to heterotic ones in \cite{Gepner:1987qi}.
   (this map is sometimes called  the ``bosonic string map" or the ``Gepner map").
      In this related heterotic ground state, there is a number of singlets under the $E_6$ group. Their number depends on the topology of the CY manifold
    and its tangent bundle, and is a useful quantity for distinguishing MIPFs.

    Using this information along with the Hodge numbers and the amount of preserved space-time supersymmetry to
    differentiate between compactification manifolds, we find that there were 421 different compactification manifolds.
     The full list of these manifolds is in Appendix~\ref{full_list}.  In addition, we find that 58 of these manifolds
     exhibit extended supersymmetry.  There is a single manifold which preserves $\N =4$ supersymmetry.  Unlike the
     case earlier, it is relatively rare that two manifolds have exactly the same Hodge numbers and number of
      singlets and yet preserve different amounts of space-time supersymmetry.  This was observed in five cases,
      which is about one percent of the total sample.

As discussed in Sect.~\ref{CFT}, there were sixty-two different model classes considered in this study.
 The 421 distinct manifolds were distributed amongst these model classes.  Ninety of these manifolds were
  found in exactly one model class.  This represents a factor of thirty increase from the earlier case of
   three manifolds being found in only one model class.  Interestingly, only three of the thirty-two
   manifolds are not represented in these ninety.  They are $\N=4~(9,9),~\N=2~(21,21)~{\rm and}~(1,1)$.
    There are again three very common manifolds.  The $\N=4~(9,9)$ is again found in every model class.
      The other two common manifolds from earlier remain very common.  This suggests that these three
      manifolds represent very symmetric cases.  This stems from the fact that they are both very common
      and the extra differentiation into singlets did not seem to have an effect upon their ubiquity.
      The general behavior for the rest of the manifolds is that they are found in a very limited number
      of different model classes with more than three quarters of all of the manifolds being found in five
       or fewer different model classes.

As the entries in Table~\ref{Hodge} suggest, this method of constructing models seems to preserve mirror
symmetry in the sense that, for each model which appears in the set, the mirror is also in the set.  However,
 we did not strictly check that each model actually has a mirror, only that another model with the correct
 Hodge numbers and the same number of singlets appeared in the set.  That is, we did not explicitly construct
a map from one model to the proposed mirror.  With that warning in mind, we shall examine the appearance
 of mirror symmetry within this set of models.

We shall start at the level of looking at the entire set of realizable models.  This is the broadest level
 possible, as we do not worry about from which model class each model comes from.  At this level,
we define the mirror of a model to be an identical model with the appropriately flipped Hodge numbers
 ({\it e.g.}~$(h_{11},h_{12}) \rightarrow(h_{12}, h_{11})$, the same amount of supersymmetry preserved,
 and the same number of singlets.  Using this definition, we find that {\it every} model has an
appropriate mirror within the model set.  This comes with the caveat that we allow for the situation
that a model is actually invariant under a mirror transformation.  That is, we allow manifolds with
$h_{12} = h_{11}$ to have odd multiplicities.  This occurs rarely, but it does occur.  In addition,
we can consider each model class separately.  Mirror symmetry even holds using this more stringent division.

In addition to considering Hodge numbers and the number of heterotic singlets, one can also differentiate
manifolds by considering how many boundary states are consistent with the model.  We shall not discuss this
 very much except to note that, the apparent mirror symmetry is broken with models where $h_{11} \ne h_{12}$.
 There exist models for which no corresponding mirror with the same number of boundary states, number
 of singlets, and appropriate Hodge numbers is in the set. This does not come as a surprise, since in the
 similar case of T-duality for circle compactifications, the number of boundary states also is not preserved
 by the duality: for radius $R^2=2N$, the T-duals have $2$ or $2N$ boundary states.

Thus far, we have only discussed what compactification manifolds we have found in our scan over the full
sixty-two model classes.  However, it is also potentially interesting to examine what is the minimum set
 of these sixty-two model classes for which every single compactification manifold is contained.
  In other words, if one just wanted to classify all of the manifolds realizable in this specific construction,
 what is the minimum set of extensions that must be considered?  Clearly, this question depends on how one
differentiates between compactification manifolds.  If we simply utilize Hodge numbers, then we would find
 every distinct manifold considering only two model classes of models.  These are, in some sense, the extreme
cases, which are every fermion paired and every fermion unpaired.  However, if we consider the singlet data
 as well as Hodge numbers then we find that although we increase the number of distinct manifolds from 32
 to 421, we only require four different model classes in order to find every manifold.  In fact, if we
only took the two model classes required in the earlier differentiation method we would only miss four manifolds.
  Thus, the two extra model classes only provide these few missing manifolds.  This is not to say that
 searching through the sixty-two different model classes for the Standard Model would be fruitless only
 that the sum total of all of the different compactification manifolds will have been found after only
 searching through these four different model classes.

We would also like to compare our results to other methods.
There are two related questions in this context.
The first is how our construction algorithm is related to the traditional fermionic construction, \cite{klt1,abk2}.
Although this search was never done to our knowledge in the IIB string we can argue that our vacua fall within the conventional definition of
fermionic constructions as these were described in \cite{klt1,abk2}.
The reason is that the simple current extension technique we use to generate MIPFs from a reference MIPF, is preserving the fermionic nature of MIPFS.
More precisely if it acts on a MIPF that is a sesquilinear form of $\theta$-functions or Ising characters, it still gives MIPFs that can be written as
    sesquilinear forms of $\theta$-functions or Ising characters.
Therefore the IIB vacua described here is a large subset of all fermionic IIB vacua.

Another set of vacua fermionic theories are usually compared to is to $\Zbf_2\times \Zbf_2$ orbifolds.
There even seems to be a ``folk theorem" stating that the two sets are equivalent\rlap.\footnote{This equivalence is between free fermionic
theories and special points in the moduli spaces of $\Zbf_2\times \Zbf_2$ orbifolds. The precise statement of such a theorem could be that
for
every free fermionic type-IIB theory there is a point in the moduli space of a type-IIB $\Zbf_2\times \Zbf_2$ orbifold matching it,   while every
type-IIB $\Zbf_2\times \Zbf_2$ orbifold has at least one point in its moduli space that can be described by free fermions.}
The second question therefore is to what extend this is true.
A first attempt was made to classify all $\Zbf_2\times \Zbf_2$ orbifolds in \cite{df}.
This task was recently completed in \cite{donagi}.
In that paper, the authors classified all  $\Zbf_2\times \Zbf_2$ orbifold actions on $(T^2)^3$ including shifts and discrete torsion.
The list of Hodge data obtained matches our table  \ref{Hodge} with one exception: our list contains in addition the Hodge numbers (25,1) and its mirror.
It is not exactly yet clear what is the origin of this mismatch. It is expected however that the simple current method is related to 
the orbifold method (or its inverse).
In particular, the (25,1) models in our list are constructed from a $\Zbf_4$ simple current extension and it is plausible that this is the reason
it is not found in \cite{donagi}.  This seems to suggest that the folk theorem stating that fermionic constructions are equivalent to $\Zbf_2\times \Zbf_2$
seems to fail. This observation requires  further study.

It is noteworthy that we have found one Hodge number pair in addition to those of $\Zbf_2\times \Zbf_2$ orbifolds, but that on the other
hand none of the Hodge number pairs of  \cite{donagi} is missing from our list.
This seems to suggest that either we are covering most, if not
all, free-fermionic theories, or the aforementioned folk theorem is not even close to being correct. To check this, it would be very
interesting to have  a complete list of Hodge numbers and heterotic singlets for all free fermionic type-IIB theories.

A partial list of closed string data for $\Zbf_2\times \Zbf_2$ orbifolds appears in \cite{Ploger:2007iq}. These authors do not
only present the Hodge numbers, but also the number of singlets and additional vector bosons in heterotic strings.  The latter
number is two in all cases, in other words the heterotic gauge group is
$E_6 \times E_8 \times U(1)^2$. Although we have not included information on the number of $U(1)$'s in this paper, we
did compute this information, so that we can compare results.
In our construction, two is the lowest number of additional gauge bosons encountered, and
it only occurs in the case of twelve unpaired free fermions (which is easily understandable, since any pairing introduces an
additional $U(1)$ factor). Of the eight spectra published in  \cite{Ploger:2007iq}, three match exactly with ours
(namely (3,51,252), (3,27,132) and (7,31,172)), whereas the five
others have a remarkably small number of singlets outside our range. In addition, four out of seven cases with $h_{11}=h_{12}$
match with ours (these were not published in \cite{Ploger:2007iq}, but communicated to us by the authors). It is not clear to us
if all of the spectra of \cite{Ploger:2007iq} can be obtained with the original free fermionic construction of \cite{abk1,klt1}, but
if they can, then this would be the first examples were the mixed fermion/boson simple current construction we use here
misses some case. 
On the other hand, 
if we impose the condition that the number of additional $U(1)$'s is exactly 2, we get a total of 83 distinct $(h_{11},h_{12}, {\rm singlets})$
cases, compared to a total of 15 in \cite{Ploger:2007iq}. There are also some differences in cases with extended supersymmetry, and
furthermore the (25,1) models absent in  \cite{donagi}  are also absent in  \cite{Ploger:2007iq}. This is not
entirely surprising, since our (25,1) spectra have 8 additional $U(1)$'s and are apparently outside the scope of \cite{Ploger:2007iq}. 
All of this just  underscores the need for a systematic comparison of
the different approaches.

\section{Standard Model Search}
One of the main goals of this study is to find a string vacuum with a low-energy limit that consists of,
 at least, a semi-realistic MSSM.  Such a vacuum could be studied from both the geometric and the
 algebraic perspectives.
In particular, its realization as a free-fermion CFT will make the process of evaluating the effective action simpler and amenable to
an algorithmic/computer treatment. This is necessary for a detailed scan of the region
 in the neighborhood of the vacuum found, as electroweak symmetry breaking, supersymmetry breaking mass generation and other important effects are
expected to be triggered by the local effective potential.

 The search methodology utilized in the present study was detailed in Ref.~\cite{adks}.
  The methodology is an implementation of the bottom-up approach \cite{akt,ibanez},  implemented in the context of
RCFT orientifolds \cite{Fuchs:2000cm} and amended with an algorithm of tadpole cancellation \cite{dhs}.
This procedure first constructs a {\bf top-down} spectrum that matches the SM by utilizing the BCFT boundary states.
This is what we call a ``top-down solution".
Such a solution can be promoted to a bona-fide string vacuum by solving the tadpole conditions.
This is achieved when possible by adding an appropriate ``hidden sector".
Since our use of the terminology ``top-down" may be confusing, let us summarize the three distinct classes of spectra that enter
the discussion. A ``bottom-up configuration" is any combination of unitary, orthogonal or symplectic gauge groups with
bi-fundamental or rank-2 tensor matter that is free of all relevant anomalies, and which might therefore be realized with a set of intersecting
branes or a set of boundary states. If such a realization is found in an explicit model, we  speak of a ``top-down solution". If in addition
a tadpole canceling hidden sector can be found (or if no hidden sector is needed to cancel all tadpoles), we call the
result a ``string vacuum".

We will describe now this procedure in a bit more detail. Full details can be found in \cite{adks} where the search criteria were developed
and where a general characterization of hypercharge embeddings was found.
The first step in the search consists of dividing the full set of boundary states (branes) present in the model into
observable and hidden sectors.  The observable sector is defined as the set of branes where Standard
Model matter resides.  This sector also gives rise to all of the Standard Model gauge symmetries.
 There are some criteria that can be placed only on the observable sector.
 These include the requirement that the $SU(3)$ and $SU(2)$ gauge symmetries
 each arise from single stacks of branes.  This eliminates the possibility that these groups arise from
 the diagonal combination of two branes.  We do not make any further assumptions about the symmetry breaking
mechanism if these gauge symmetries are embedded in larger groups.  Hypercharge is allowed to arise
from any massless linear combination of $U(1)$ factors arising from observable sector stacks of branes.
Next we require that there be the matter content consistent
 with the three generation MSSM present in the observable sector, and no chiral exotics (see below). Furthermore
 we require that the observable sector consists of no more than four distinct stacks of branes, in order to
keep the search manageable. With more stacks of branes, the number of ways of embedding the hypercharge $Y$
increases drastically, and one may also obtain quarks and leptons from several distinct bi-fundamentals. On the other
hand, the number of options for chiral exotics increases. It is not clear which of these competing effects dominates.

  As our definition of what constitutes a chiral exotic may differ a bit from that usually used in the
literature, we shall now define chiral exotics for these models.  We do not put any constraints on
 matter that is not charged under the observable Chan-Paton group, {\it i.e.}
 we allow for any amount of chiral matter that is limited to the hidden sector.
 If the chiral matter is in
the observable sector we require it either be part of the MSSM spectrum or at least non-chiral with
 respect to all of the Standard Model gauge groups.
Apart from the standard three families of quarks, charged leptons and left-handed neutrinos, this definition
does allow a few more particles that are chiral with respect to the observable part of the Chan-Paton
group\rlap.\footnote{The precise definition of the ``observable part of the Chan-Paton group"
is those factors of the original Chan-Paton group that contain parts of $SU(3)\times SU(2)\times U(1)$,
before taking breaking through axion mixing into account.}
 It allows for right-handed neutrinos that are chiral with respect to an extension of the standard model
(the most common case being a broken or unbroken gauged $B-L$ symmetry).
 It also allows Higgs pair candidates
 that are chiral with respect to a $U(2)$ group, which contains $SU(2)_{\rm Weak}$ (in this case the additional
 $U(1)$ is broken by axion mixing). Among the less desirable particles in this category
 are mirror pairs of quarks and leptons that are chiral with respect to the Chan-Paton group, but
 non-chiral with respect to the standard model gauge group. Note that although the latter particles are exotic
 and chiral with respect to the full Chan-Paton group, we do not call them ``chiral exotics"  because they are not
 chiral with respect to $SU(3)\times SU(2)\times U(1)$.

 Apart from chiral observable and chiral hidden matter, a third category is chiral observable-hidden matter.
 Such matter may be subject to symmetry breaking or confinement in the hidden sector, and is therefore
 not necessarily fatal. Furthermore, there are several kinds of chiral observable-hidden matter that nevertheless
 fulfill the requirements stated in the previous paragraph, {\it i.e.} that they are non-chiral with respect to the standard model
 gauge group.
 Nevertheless, in the previous searches \cite{dhs} and \cite{adks} chiral observable-hidden matter
 was not accepted. In other words, boundary states with a chiral intersection with the standard model
 branes were given Chan-Paton multiplicity zero. This has the advantage of limiting the scope
 of the search to the {\it a priori} most attractive models. In a few cases where this requirement was lifted, this
 resulted in an explosion of the number of solutions by several orders of magnitude.
 In situations where the main search result is negative, it is natural to remove this requirement. This
 is indeed what we have done in the present paper.

Because of  the existence of several possible definitions of chiral exotics, we wish to emphasize that
 most spectra obtained in previous searches are free
of chiral exotics, for {\it any} definition of the latter. For example, apart from
 three chiral right-handed neutrinos that are chiral with respect to $B-L$,  but not ``exotic" by any standard,
about $85\%$ of the  about 200.000 spectra collected
 in \cite{dhs} have
  have no extra chiral matter at all, $12.5\%$ has a chiral hidden sector, and about
 $2\%$ have a $U(2)$ chiral Higgs pair and/or chiral mirror pairs.

Utilizing the criteria outlined above, we found that only 1 of the 62 model classes yielded any top-down solutions.
This model class was the case of all fermions paired (that is,
 simple current extensions only involving powers of $D_1$), or in other words a compactification that can
 be realized entirely using free bosons and self-dual lattices \cite{LLS}.
 No other model classes yielded models that satisfied these criteria. The search was done without
 any constraint on the number of boundary states. In  \cite{adks} an upper limit of 1750 was used. In the present case
 the number of boundary states goes up to 3040, but in most cases already the first step in the search (looking for
 three quark doublets) failed. Thus because of lack of results, larger numbers became accessible.

 We did have to impose a limitation on the scope of the MIPF search. The most difficult case, twelve real and three complex
 fermions, has 534700 MIPFs. As explained earlier, not all of these are distinct. The vast majority of this large number
 comes from the discrete torsion signs of large simple current subgroups.
 Since the simple current group in this case
 is $({\bf Z}_2)^7$, the largest subgroup, the simple current group itself, admits 21 such signs (they form an anti-symmetric
 $7 \times 7$ matrix \cite{Kreuzer:1993tf}). This leads to $2^{21}$ possibilities, still subject to identification by permutations.
 It turns out that these in principle distinct MIPFs produce very few distinct Hodge numbers. For this reason we have searched
 the MIPFs originating from large simple current subgroups by taking a random sample of 100 discrete torsion sign choices per subgroup\rlap.\footnote{This kind of sampling was only done for the Standard Model search. The Hodge number scan was done completely, and
 gave rise to many degeneracies for a given simple current subgroup.}

 The top-down solutions we found were of a chiral type already encountered in \cite{adks} for Gepner models.
 The simplest of them is a Pati-Salam type of spectrum, which is remarkably simple. The Chan-Paton group is
 $U(4)\times U(2) \times U(2)$, with all $U(1)$ symmetries broken by axion mixing
 (note that $Y$ is  the $U(4)$ generator $\frac{1}{6}(1,1,1,-3)$). The spectrum consists
 of the following left-handed particles (with ``$V$" for vector and ``$V^*$" for conjugate vector)
 \begin{eqnarray*}
 2 \  \times &(V,V,0) \\   &(V,V^*,0) \\ 2\ \times  &(V^*,0,V^*) \\  &(V^*,0,V) \\   2\ \times  &(0,V,V^*) \\
 \end{eqnarray*}
 which represent respectively three $SU(4)$-unified quark and lepton doublets, three $SU(4)$ unified
 anti-quark and charged lepton singlets, and 2 particles with the quantum numbers of a MSSM Higgs pair.
 Therefore, apart from the $U(4)$ baryon-lepton unification and the extra Higgs pair this is precisely the MSSM spectrum.
 We emphasize that the multiplicities given above are the exact multiplicities of left-handed particles, and not the
 net number (left minus right). Hence the additional Higgs pair is the only exotic, there are no mirror quarks or leptons
 whatsoever, not even fully non-chiral ones. This is extremely rare, and we do no know any such example in the entire set
 of spectra obtained from Gepner models\footnote{Note however that the available spectra in the Gepner model search, \cite{dhs},
 are free of tadpoles; there is no databases of exact spectra of top-down solutions prior to tadpole cancellation.
 The search performed in \cite{adks} focused more on chiral types than on tadpole solutions,
 but the chiral types were collected modulo non-chiral exotics, so that there is no such database in that case either.}

 The second chiral type we found is essentially the same as the foregoing, but with the $SU(4)$ stack split in three
 baryon and one lepton stack. This spectrum has one additional exotic, a non-chiral set of leptoquarks
 originating from the gaugino corresponding to the
 broken generators of $SU(4)$.

 However, even after relaxing the observable-hidden chirality constraint, as explained above, we were unable
 to obtain a solution to the tadpole conditions for any of these models.

  As an extra check on the top-down model search algorithm, we relaxed the requirement that there be exactly three
 generations and found numerous examples of one and two generation models in many different model classes.  We
 tried this on a total of 65 MIPFs, a small fraction of the total, and found top-down configurations in 62 of them.
 Tadpole solution were found for some one-family models, but not for two-family models.
Due to the limited number of cases considered, no conclusions with regard to
 family statistics should be drawn from these observations.
 But this does reinforce the finding that there are only very few models with three generations in the entire
 set of models constructed. Despite the need for statistical sampling mentioned above, it seems extremely
 unlikely to us that any three family models were missed.

\vskip 2cm
\centerline{\bf Acknowledgements}
\addcontentsline{toc}{section}{Acknowledgements}
\vskip 2cm

We would like to thank Massimo Bianchi and Ron Donagi for discussions, and
Patrick Vaudrevange for bringing  \cite{Ploger:2007iq} to our attention, and providing
some additional information. 
This work was partially supported by ANR grant NT05-1-41861,
RTN contracts MRTN-CT-2004-005104 and MRTN-CT-2004-503369,
CNRS PICS  3059 and 3747 and by a European Excellence Grant,
MEXT-CT-2003-509661.The work of A.N.S. has been performed as part of the program
FP 57 of the Foundation for Fundamental Research of Matter (FOM), and has
been partially
supported by funding of the Spanish ``Ministerio de
Ciencia y Tecnolog\'\i a", Project BFM2002-03610.
Elias Kiritsis is on leave from CPHT, Ecole Polytechnique (UMR du CNRS 7644).


\appendix

\vskip 10mm
\renewcommand{\theequation}{\thesection.\arabic{equation}}
\centerline{\Large\bf Appendices}
\addcontentsline{toc}{section}{Appendices}
\section{Full List of Manifolds}

The following table contains the full list of compactification manifolds found during this search.
We have organized the table in such a way as to include the Hodge numbers the number of $E_6$
singlets (listed as heterotic singlets on the table), the amount of space-time supersymmetry
 preserved, and the number of boundary states.  The boundary state information includes the
 following values:  the maximum value that the number of boundary states took, the minimum value
 for the number of boundary states, and the total number of different values for the number of
 boundary states.  For a more complete discussion of all of this information see Ref.~\cite{dhs}.

\label{full_list}

\begin{center}
\begin{longtable}{||c|c|c|c|c|c||}
\hline
\hline
\multicolumn{1}{||l|}{Hodge numbers}
& \multicolumn{1}{c|}{Heterotic}
& \multicolumn{1}{c|}{$d=4$}
& \multicolumn{3}{c||}{Boundary States}\\
\multicolumn{1}{||c|}{$(h_{11},h_{12})$}
& \multicolumn{1}{c|}{Singlets}
& \multicolumn{1}{c|}{Susy}
& \multicolumn{1}{c}{Maximum}
& \multicolumn{1}{c}{Minimum}
& \multicolumn{1}{c||}{Distinct}\\
\hline
\endfirsthead
\multicolumn{6}{r}%
{{\bfseries 
{\rm-- continued from previous page}}} \\  \hline
\multicolumn{1}{||l|}{Hodge numbers}
& \multicolumn{1}{c|}{Heterotic}
& \multicolumn{1}{c|}{$d=4$}
& \multicolumn{3}{c||}{Boundary States}\\
\multicolumn{1}{||c|}{$(h_{11},h_{12})$}
& \multicolumn{1}{c|}{Singlets}
& \multicolumn{1}{c|}{Susy}
& \multicolumn{1}{c}{Maximum}
& \multicolumn{1}{c}{Minimum}
& \multicolumn{1}{c||}{Distinct}\\
\hline
\endhead
\hline \multicolumn{6}{||r||}{{continued on next page}} \\ \hline
\endfoot
\hline \hline
\endlastfoot
$(51,3)$ &  258 &  $ \N = 1$ &  2048 &  32 & 14\\
$(51,3)$ &  256 &  $ \N = 1$ &  2272 &  160 & 16\\
$(51,3)$ &  254 &  $ \N = 1$ &  2528 &  448 & 12\\
$(51,3)$ &  252 &  $ \N = 1$ &  3040 &  2048 & 5\\
$(3,51)$ &  258 &  $ \N = 1$ &  1280 &  32 & 11\\
$(3,51)$ &  256 &  $ \N = 1$ &  1504 &  128 & 15\\
$(3,51)$ &  254 &  $ \N = 1$ &  1760 &  416 & 12\\
$(3,51)$ &  252 &  $ \N = 1$ &  2272 &  1280 & 5\\
\hline
$(31,7)$ &  254 &  $ \N = 1$ &  1216 &  32 & 20\\
$(31,7)$ &  252 &  $ \N = 1$ &  1376 &  128 & 21\\
$(31,7)$ &  230 &  $ \N = 1$ &  1376 &  160 & 18\\
$(31,7)$ &  228 &  $ \N = 1$ &  1552 &  496 & 15\\
$(31,7)$ &  209 &  $ \N = 1$ &  1600 &  152 & 23\\
$(31,7)$ &  208 &  $ \N = 1$ &  1376 &  128 & 22\\
$(31,7)$ &  207 &  $ \N = 1$ &  1952 &  592 & 15\\
$(31,7)$ &  206 &  $ \N = 1$ &  1552 &  416 & 17\\
$(31,7)$ &  190 &  $ \N = 1$ &  1600 &  320 & 16\\
$(31,7)$ &  188 &  $ \N = 1$ &  1952 &  1312 & 10\\
$(31,7)$ &  174 &  $ \N = 1$ &  1600 &  256 & 22\\
$(31,7)$ &  172 &  $ \N = 1$ &  1952 &  1088 & 12\\
$(7,31)$ &  254 &  $ \N = 1$ &  704 &  32 & 14\\
$(7,31)$ &  252 &  $ \N = 1$ &  992 &  128 & 13\\
$(7,31)$ &  230 &  $ \N = 1$ &  992 &  128 & 16\\
$(7,31)$ &  228 &  $ \N = 1$ &  1168 &  416 & 11\\
$(7,31)$ &  209 &  $ \N = 1$ &  1216 &  152 & 21\\
$(7,31)$ &  208 &  $ \N = 1$ &  992 &  64 & 18\\
$(7,31)$ &  207 &  $ \N = 1$ &  1568 &  592 & 13\\
$(7,31)$ &  206 &  $ \N = 1$ &  1168 &  256 & 14\\
$(7,31)$ &  190 &  $ \N = 1$ &  1216 &  304 & 14\\
$(7,31)$ &  188 &  $ \N = 1$ &  1568 &  928 & 10\\
$(7,31)$ &  174 &  $ \N = 1$ &  1216 &  224 & 18\\
$(7,31)$ &  172 &  $ \N = 1$ &  1568 &  704 & 12\\
\hline
$(27,3)$ &  270 &  $ \N = 1$ &  448 &  8 & 16\\
$(27,3)$ &  240 &  $ \N = 1$ &  1024 &  40 & 16\\
$(27,3)$ &  234 &  $ \N = 1$ &  1024 &  32 & 15\\
$(27,3)$ &  216 &  $ \N = 1$ &  1184 &  128 & 16\\
$(27,3)$ &  213 &  $ \N = 1$ &  1184 &  110 & 18\\
$(27,3)$ &  212 &  $ \N = 1$ &  1024 &  128 & 12\\
$(27,3)$ &  200 &  $ \N = 1$ &  1184 &  128 & 21\\
$(27,3)$ &  198 &  $ \N = 1$ &  1360 &  392 & 13\\
$(27,3)$ &  189 &  $ \N = 1$ &  1504 &  172 & 18\\
$(27,3)$ &  188 &  $ \N = 1$ &  1184 &  496 & 8\\
$(27,3)$ &  182 &  $ \N = 1$ &  1360 &  304 & 23\\
$(27,3)$ &  180 &  $ \N = 1$ &  1664 &  1312 & 6\\
$(27,3)$ &  167 &  $ \N = 1$ &  1504 &  440 & 16\\
$(27,3)$ &  166 &  $ \N = 1$ &  1360 &  224 & 27\\
$(27,3)$ &  164 &  $ \N = 1$ &  1664 &  1168 & 11\\
$(27,3)$ &  148 &  $ \N = 1$ &  1664 &  992 & 15\\
$(27,3)$ &  132 &  $ \N = 1$ &  1664 &  896 & 16\\
$(3,27)$ &  270 &  $ \N = 1$ &  256 &  8 & 12\\
$(3,27)$ &  240 &  $ \N = 1$ &  608 &  32 & 14\\
$(3,27)$ &  234 &  $ \N = 1$ &  448 &  32 & 9\\
$(3,27)$ &  216 &  $ \N = 1$ &  800 &  128 & 11\\
$(3,27)$ &  213 &  $ \N = 1$ &  800 &  86 & 18\\
$(3,27)$ &  212 &  $ \N = 1$ &  320 &  64 & 8\\
$(3,27)$ &  200 &  $ \N = 1$ &  800 &  64 & 12\\
$(3,27)$ &  198 &  $ \N = 1$ &  976 &  392 & 10\\
$(3,27)$ &  189 &  $ \N = 1$ &  1120 &  172 & 18\\
$(3,27)$ &  188 &  $ \N = 1$ &  800 &  304 & 7\\
$(3,27)$ &  182 &  $ \N = 1$ &  976 &  256 & 17\\
$(3,27)$ &  180 &  $ \N = 1$ &  1280 &  928 & 6\\
$(3,27)$ &  167 &  $ \N = 1$ &  1120 &  344 & 16\\
$(3,27)$ &  166 &  $ \N = 1$ &  976 &  128 & 18\\
$(3,27)$ &  164 &  $ \N = 1$ &  1280 &  784 & 11\\
$(3,27)$ &  148 &  $ \N = 1$ &  1280 &  608 & 15\\
$(3,27)$ &  132 &  $ \N = 1$ &  1280 &  608 & 15\\
\hline
$(25,1)$ &  230 &  $ \N = 1$ &  256 &  32 & 4\\
$(1,25)$ &  230 &  $ \N = 1$ &  64 &  32 & 2\\
\hline
$(21,9)$ &  172 &  $ \N = 1$ &  1184 &  64 & 26\\
$(21,9)$ &  170 &  $ \N = 1$ &  1504 &  160 & 19\\
$(21,9)$ &  169 &  $ \N = 1$ &  1120 &  152 & 22\\
$(21,9)$ &  167 &  $ \N = 1$ &  1312 &  496 & 13\\
$(21,9)$ &  166 &  $ \N = 1$ &  1184 &  304 & 12\\
$(21,9)$ &  164 &  $ \N = 1$ &  1504 &  1088 & 8\\
$(9,21)$ &  172 &  $ \N = 1$ &  992 &  32 & 23\\
$(9,21)$ &  170 &  $ \N = 1$ &  1312 &  160 & 18\\
$(9,21)$ &  169 &  $ \N = 1$ &  928 &  124 & 21\\
$(9,21)$ &  167 &  $ \N = 1$ &  1120 &  400 & 13\\
$(9,21)$ &  166 &  $ \N = 1$ &  992 &  296 & 10\\
$(9,21)$ &  164 &  $ \N = 1$ &  1312 &  896 & 8\\
\hline
$(19,7)$ &  208 &  $ \N = 1$ &  832 &  128 & 17\\
$(19,7)$ &  202 &  $ \N = 1$ &  832 &  128 & 20\\
$(19,7)$ &  196 &  $ \N = 1$ &  832 &  128 & 13\\
$(19,7)$ &  187 &  $ \N = 1$ &  992 &  172 & 18\\
$(19,7)$ &  184 &  $ \N = 1$ &  992 &  392 & 14\\
$(19,7)$ &  181 &  $ \N = 1$ &  992 &  152 & 21\\
$(19,7)$ &  178 &  $ \N = 1$ &  992 &  392 & 11\\
$(19,7)$ &  168 &  $ \N = 1$ &  992 &  304 & 20\\
$(19,7)$ &  166 &  $ \N = 1$ &  1264 &  196 & 21\\
$(19,7)$ &  163 &  $ \N = 1$ &  1264 &  448 & 15\\
$(19,7)$ &  162 &  $ \N = 1$ &  992 &  304 & 18\\
$(19,7)$ &  160 &  $ \N = 1$ &  1264 &  1072 & 5\\
$(19,7)$ &  147 &  $ \N = 1$ &  1264 &  392 & 23\\
$(19,7)$ &  144 &  $ \N = 1$ &  1264 &  896 & 9\\
$(19,7)$ &  128 &  $ \N = 1$ &  1264 &  736 & 12\\
$(7,19)$ &  208 &  $ \N = 1$ &  320 &  128 & 6\\
$(7,19)$ &  202 &  $ \N = 1$ &  608 &  64 & 14\\
$(7,19)$ &  196 &  $ \N = 1$ &  608 &  64 & 10\\
$(7,19)$ &  187 &  $ \N = 1$ &  800 &  172 & 10\\
$(7,19)$ &  184 &  $ \N = 1$ &  800 &  304 & 11\\
$(7,19)$ &  181 &  $ \N = 1$ &  800 &  152 & 17\\
$(7,19)$ &  178 &  $ \N = 1$ &  800 &  304 & 9\\
$(7,19)$ &  168 &  $ \N = 1$ &  800 &  224 & 15\\
$(7,19)$ &  166 &  $ \N = 1$ &  1072 &  196 & 16\\
$(7,19)$ &  163 &  $ \N = 1$ &  1072 &  392 & 13\\
$(7,19)$ &  162 &  $ \N = 1$ &  800 &  224 & 14\\
$(7,19)$ &  160 &  $ \N = 1$ &  1072 &  880 & 5\\
$(7,19)$ &  147 &  $ \N = 1$ &  1072 &  304 & 20\\
$(7,19)$ &  144 &  $ \N = 1$ &  1072 &  704 & 9\\
$(7,19)$ &  128 &  $ \N = 1$ &  1072 &  544 & 12\\
\hline
$(17,5)$ &  238 &  $ \N = 1$ &  128 &  32 & 3\\
$(17,5)$ &  176 &  $ \N = 1$ &  832 &  32 & 25\\
$(17,5)$ &  173 &  $ \N = 1$ &  800 &  86 & 18\\
$(17,5)$ &  167 &  $ \N = 1$ &  800 &  172 & 10\\
$(17,5)$ &  164 &  $ \N = 1$ &  832 &  32 & 25\\
$(17,5)$ &  161 &  $ \N = 1$ &  896 &  124 & 33\\
$(17,5)$ &  158 &  $ \N = 1$ &  896 &  296 & 18\\
$(17,5)$ &  152 &  $ \N = 1$ &  1088 &  64 & 30\\
$(17,5)$ &  149 &  $ \N = 1$ &  992 &  152 & 25\\
$(17,5)$ &  146 &  $ \N = 1$ &  1088 &  160 & 31\\
$(17,5)$ &  143 &  $ \N = 1$ &  1120 &  344 & 27\\
$(17,5)$ &  140 &  $ \N = 1$ &  1120 &  896 & 7\\
$(17,5)$ &  130 &  $ \N = 1$ &  1088 &  128 & 24\\
$(17,5)$ &  127 &  $ \N = 1$ &  1120 &  304 & 31\\
$(17,5)$ &  124 &  $ \N = 1$ &  1120 &  736 & 12\\
$(5,17)$ &  238 &  $ \N = 1$ &  128 &  32 & 3\\
$(5,17)$ &  176 &  $ \N = 1$ &  640 &  32 & 15\\
$(5,17)$ &  173 &  $ \N = 1$ &  608 &  62 & 17\\
$(5,17)$ &  167 &  $ \N = 1$ &  608 &  124 & 9\\
$(5,17)$ &  164 &  $ \N = 1$ &  640 &  32 & 19\\
$(5,17)$ &  161 &  $ \N = 1$ &  704 &  124 & 20\\
$(5,17)$ &  158 &  $ \N = 1$ &  704 &  224 & 14\\
$(5,17)$ &  152 &  $ \N = 1$ &  896 &  64 & 25\\
$(5,17)$ &  149 &  $ \N = 1$ &  800 &  152 & 21\\
$(5,17)$ &  146 &  $ \N = 1$ &  896 &  128 & 23\\
$(5,17)$ &  143 &  $ \N = 1$ &  928 &  304 & 20\\
$(5,17)$ &  140 &  $ \N = 1$ &  928 &  704 & 7\\
$(5,17)$ &  130 &  $ \N = 1$ &  896 &  64 & 20\\
$(5,17)$ &  127 &  $ \N = 1$ &  928 &  248 & 26\\
$(5,17)$ &  124 &  $ \N = 1$ &  928 &  544 & 12\\
\hline
$(15,3)$ &  222 &  $ \N = 1$ &  64 &  32 & 2\\
$(15,3)$ &  160 &  $ \N = 1$ &  160 &  32 & 4\\
$(15,3)$ &  138 &  $ \N = 1$ &  928 &  16 & 19\\
$(15,3)$ &  132 &  $ \N = 1$ &  832 &  64 & 23\\
$(15,3)$ &  129 &  $ \N = 1$ &  928 &  124 & 33\\
$(15,3)$ &  126 &  $ \N = 1$ &  928 &  128 & 24\\
$(15,3)$ &  123 &  $ \N = 1$ &  928 &  304 & 23\\
$(15,3)$ &  120 &  $ \N = 1$ &  976 &  736 & 8\\
$(3,15)$ &  222 &  $ \N = 1$ &  64 &  32 & 2\\
$(3,15)$ &  160 &  $ \N = 1$ &  64 &  32 & 2\\
$(3,15)$ &  138 &  $ \N = 1$ &  416 &  16 & 13\\
$(3,15)$ &  132 &  $ \N = 1$ &  608 &  32 & 17\\
$(3,15)$ &  129 &  $ \N = 1$ &  736 &  124 & 23\\
$(3,15)$ &  126 &  $ \N = 1$ &  736 &  64 & 17\\
$(3,15)$ &  123 &  $ \N = 1$ &  736 &  248 & 17\\
$(3,15)$ &  120 &  $ \N = 1$ &  784 &  544 & 8\\
\hline
$(12,6)$ &  129 &  $ \N = 1$ &  848 &  62 & 24\\
$(12,6)$ &  126 &  $ \N = 1$ &  848 &  148 & 30\\
$(12,6)$ &  123 &  $ \N = 1$ &  848 &  304 & 18\\
$(12,6)$ &  120 &  $ \N = 1$ &  848 &  688 & 6\\
$(6,12)$ &  129 &  $ \N = 1$ &  752 &  62 & 20\\
$(6,12)$ &  126 &  $ \N = 1$ &  752 &  112 & 24\\
$(6,12)$ &  123 &  $ \N = 1$ &  752 &  272 & 15\\
$(6,12)$ &  120 &  $ \N = 1$ &  752 &  592 & 6\\
\hline
$(21,21)$ &  160 &  $ \N = 2$ &  1600 &  16 & 15\\
$(21,21)$ &  148 &  $ \N = 2$ &  1760 &  80 & 16\\
$(21,21)$ &  144 &  $ \N = 2$ &  2240 &  16 & 31\\
$(21,21)$ &  140 &  $ \N = 2$ &  2560 &  64 & 43\\
$(21,21)$ &  136 &  $ \N = 2$ &  3392 &  256 & 34\\
\hline
$(19,19)$ &  242 &  $ \N = 1$ &  1280 &  32 & 15\\
$(19,19)$ &  240 &  $ \N = 1$ &  1504 &  128 & 22\\
$(19,19)$ &  238 &  $ \N = 1$ &  1760 &  416 & 14\\
$(19,19)$ &  208 &  $ \N = 1$ &  1504 &  128 & 14\\
$(19,19)$ &  206 &  $ \N = 1$ &  1760 &  448 & 13\\
$(19,19)$ &  204 &  $ \N = 1$ &  2272 &  1472 & 7\\
$(19,19)$ &  180 &  $ \N = 1$ &  1472 &  128 & 14\\
$(19,19)$ &  178 &  $ \N = 1$ &  1696 &  320 & 12\\
$(19,19)$ &  176 &  $ \N = 1$ &  1952 &  128 & 21\\
$(19,19)$ &  174 &  $ \N = 1$ &  1760 &  416 & 12\\
$(19,19)$ &  172 &  $ \N = 1$ &  2272 &  1280 & 5\\
\hline
$(15,15)$ &  270 &  $ \N = 1$ &  448 &  32 & 16\\
$(15,15)$ &  240 &  $ \N = 1$ &  832 &  128 & 18\\
$(15,15)$ &  234 &  $ \N = 1$ &  448 &  32 & 9\\
$(15,15)$ &  216 &  $ \N = 1$ &  992 &  128 & 15\\
$(15,15)$ &  213 &  $ \N = 1$ &  992 &  172 & 22\\
$(15,15)$ &  212 &  $ \N = 1$ &  832 &  64 & 16\\
$(15,15)$ &  200 &  $ \N = 1$ &  1024 &  64 & 22\\
$(15,15)$ &  198 &  $ \N = 1$ &  1184 &  392 & 16\\
$(15,15)$ &  190 &  $ \N = 1$ &  992 &  128 & 16\\
$(15,15)$ &  189 &  $ \N = 1$ &  1312 &  536 & 10\\
$(15,15)$ &  188 &  $ \N = 1$ &  1168 &  304 & 19\\
$(15,15)$ &  184 &  $ \N = 1$ &  1024 &  64 & 15\\
$(15,15)$ &  182 &  $ \N = 1$ &  1216 &  224 & 28\\
$(15,15)$ &  180 &  $ \N = 1$ &  1568 &  1120 & 10\\
$(15,15)$ &  169 &  $ \N = 1$ &  1216 &  152 & 23\\
$(15,15)$ &  167 &  $ \N = 1$ &  1568 &  440 & 21\\
$(15,15)$ &  166 &  $ \N = 1$ &  1216 &  224 & 29\\
$(15,15)$ &  164 &  $ \N = 1$ &  1568 &  896 & 17\\
$(15,15)$ &  160 &  $ \N = 1$ &  1184 &  64 & 25\\
$(15,15)$ &  158 &  $ \N = 1$ &  1360 &  208 & 20\\
$(15,15)$ &  150 &  $ \N = 1$ &  1216 &  304 & 14\\
$(15,15)$ &  148 &  $ \N = 1$ &  1568 &  800 & 18\\
$(15,15)$ &  138 &  $ \N = 1$ &  1184 &  256 & 13\\
$(15,15)$ &  136 &  $ \N = 1$ &  1360 &  896 & 6\\
$(15,15)$ &  132 &  $ \N = 1$ &  1472 &  704 & 16\\
\hline
$(13,13)$ &  230 &  $ \N = 1$ &  256 &  32 & 4\\
$(13,13)$ &  192 &  $ \N = 2$ &  320 &  8 & 10\\
$(13,13)$ &  172 &  $ \N = 2$ &  704 &  40 & 9\\
$(13,13)$ &  160 &  $ \N = 2$ &  896 &  8 & 22\\
$(13,13)$ &  156 &  $ \N = 2$ &  1024 &  64 & 29\\
$(13,13)$ &  148 &  $ \N = 2$ &  1024 &  64 & 24\\
$(13,13)$ &  144 &  $ \N = 2$ &  1184 &  8 & 28\\
$(13,13)$ &  140 &  $ \N = 2$ &  1024 &  32 & 28\\
$(13,13)$ &  136 &  $ \N = 2$ &  1184 &  256 & 17\\
$(13,13)$ &  128 &  $ \N = 2$ &  1024 &  80 & 14\\
$(13,13)$ &  120 &  $ \N = 2$ &  1216 &  16 & 31\\
$(13,13)$ &  116 &  $ \N = 2$ &  1472 &  128 & 31\\
$(13,13)$ &  112 &  $ \N = 2$ &  1472 &  32 & 36\\
$(13,13)$ &  108 &  $ \N = 2$ &  1984 &  128 & 31\\
$(13,13)$ &  104 &  $ \N = 2$ &  1984 &  896 & 12\\
$(13,13)$ &  96 &  $ \N = 2$ &  448 &  32 & 11\\
$(13,13)$ &  84 &  $ \N = 2$ &  1024 &  128 & 19\\
$(13,13)$ &  80 &  $ \N = 2$ &  1216 &  32 & 18\\
$(13,13)$ &  76 &  $ \N = 2$ &  1472 &  128 & 28\\
$(13,13)$ &  72 &  $ \N = 2$ &  1984 &  608 & 18\\
\hline
$(11,11)$ &  238 &  $ \N = 1$ &  704 &  32 & 14\\
$(11,11)$ &  226 &  $ \N = 1$ &  896 &  32 & 12\\
$(11,11)$ &  224 &  $ \N = 1$ &  1120 &  128 & 16\\
$(11,11)$ &  214 &  $ \N = 1$ &  832 &  128 & 19\\
$(11,11)$ &  204 &  $ \N = 1$ &  832 &  128 & 16\\
$(11,11)$ &  200 &  $ \N = 1$ &  704 &  32 & 17\\
$(11,11)$ &  193 &  $ \N = 1$ &  1024 &  172 & 21\\
$(11,11)$ &  192 &  $ \N = 1$ &  1120 &  64 & 23\\
$(11,11)$ &  190 &  $ \N = 1$ &  1376 &  416 & 18\\
$(11,11)$ &  180 &  $ \N = 1$ &  992 &  392 & 17\\
$(11,11)$ &  176 &  $ \N = 1$ &  832 &  64 & 29\\
$(11,11)$ &  174 &  $ \N = 1$ &  1024 &  304 & 20\\
$(11,11)$ &  173 &  $ \N = 1$ &  800 &  86 & 26\\
$(11,11)$ &  172 &  $ \N = 1$ &  800 &  64 & 13\\
$(11,11)$ &  170 &  $ \N = 1$ &  832 &  32 & 25\\
$(11,11)$ &  167 &  $ \N = 1$ &  704 &  124 & 18\\
$(11,11)$ &  164 &  $ \N = 1$ &  1088 &  64 & 29\\
$(11,11)$ &  162 &  $ \N = 1$ &  1312 &  320 & 13\\
$(11,11)$ &  161 &  $ \N = 1$ &  800 &  124 & 25\\
$(11,11)$ &  160 &  $ \N = 1$ &  1120 &  64 & 19\\
$(11,11)$ &  159 &  $ \N = 1$ &  1376 &  440 & 21\\
$(11,11)$ &  158 &  $ \N = 1$ &  1376 &  224 & 33\\
$(11,11)$ &  156 &  $ \N = 1$ &  1888 &  896 & 18\\
$(11,11)$ &  152 &  $ \N = 1$ &  992 &  160 & 22\\
$(11,11)$ &  149 &  $ \N = 1$ &  1120 &  152 & 24\\
$(11,11)$ &  148 &  $ \N = 1$ &  976 &  304 & 15\\
$(11,11)$ &  146 &  $ \N = 1$ &  992 &  152 & 29\\
$(11,11)$ &  143 &  $ \N = 1$ &  1024 &  86 & 40\\
$(11,11)$ &  142 &  $ \N = 1$ &  1024 &  160 & 29\\
$(11,11)$ &  140 &  $ \N = 1$ &  1376 &  784 & 18\\
$(11,11)$ &  131 &  $ \N = 1$ &  1216 &  440 & 14\\
$(11,11)$ &  130 &  $ \N = 1$ &  1312 &  128 & 27\\
$(11,11)$ &  129 &  $ \N = 1$ &  1024 &  124 & 18\\
$(11,11)$ &  128 &  $ \N = 1$ &  1568 &  896 & 12\\
$(11,11)$ &  127 &  $ \N = 1$ &  1376 &  296 & 35\\
$(11,11)$ &  126 &  $ \N = 1$ &  1376 &  224 & 23\\
$(11,11)$ &  124 &  $ \N = 1$ &  1888 &  608 & 27\\
$(11,11)$ &  120 &  $ \N = 1$ &  992 &  64 & 19\\
$(11,11)$ &  118 &  $ \N = 1$ &  1168 &  152 & 28\\
$(11,11)$ &  114 &  $ \N = 1$ &  1024 &  16 & 21\\
$(11,11)$ &  112 &  $ \N = 1$ &  1216 &  64 & 32\\
$(11,11)$ &  108 &  $ \N = 1$ &  1376 &  608 & 18\\
$(11,11)$ &  102 &  $ \N = 1$ &  1280 &  256 & 10\\
$(11,11)$ &  100 &  $ \N = 1$ &  1504 &  896 & 5\\
$(11,11)$ &  98 &  $ \N = 1$ &  1312 &  256 & 11\\
$(11,11)$ &  96 &  $ \N = 1$ &  1568 &  704 & 13\\
$(11,11)$ &  94 &  $ \N = 1$ &  1376 &  224 & 11\\
$(11,11)$ &  92 &  $ \N = 1$ &  1888 &  800 & 11\\
\hline
$(9,9)$ &  222 &  $ \N = 1$ &  128 &  32 & 3\\
$(9,9)$ &  160 &  $ \N = 1$ &  736 &  296 & 14\\
$(9,9)$ &  154 &  $ \N = 1$ &  736 &  296 & 12\\
$(9,9)$ &  147 &  $ \N = 1$ &  704 &  124 & 18\\
$(9,9)$ &  144 &  $ \N = 2$ &  128 &  32 & 3\\
$(9,9)$ &  144 &  $ \N = 1$ &  704 &  296 & 11\\
$(9,9)$ &  141 &  $ \N = 1$ &  704 &  124 & 17\\
$(9,9)$ &  139 &  $ \N = 1$ &  976 &  368 & 14\\
$(9,9)$ &  138 &  $ \N = 1$ &  832 &  64 & 30\\
$(9,9)$ &  136 &  $ \N = 1$ &  976 &  784 & 6\\
$(9,9)$ &  132 &  $ \N = 1$ &  736 &  32 & 24\\
$(9,9)$ &  129 &  $ \N = 1$ &  832 &  124 & 29\\
$(9,9)$ &  126 &  $ \N = 1$ &  976 &  128 & 35\\
$(9,9)$ &  123 &  $ \N = 1$ &  976 &  272 & 27\\
$(9,9)$ &  120 &  $ \N = 1$ &  976 &  608 & 11\\
$(9,9)$ &  117 &  $ \N = 1$ &  896 &  86 & 16\\
$(9,9)$ &  114 &  $ \N = 1$ &  896 &  152 & 21\\
$(9,9)$ &  111 &  $ \N = 1$ &  896 &  392 & 9\\
$(9,9)$ &  108 &  $ \N = 1$ &  896 &  800 & 4\\
$(9,9)$ &  107 &  $ \N = 1$ &  976 &  296 & 20\\
$(9,9)$ &  104 &  $ \N = 1$ &  976 &  608 & 9\\
$(9,9)$ &  100 &  $ \N = 2$ &  352 &  40 & 8\\
$(9,9)$ &  98 &  $ \N = 1$ &  896 &  112 & 32\\
$(9,9)$ &  96 &  $ \N = 2$ &  896 &  16 & 26\\
$(9,9)$ &  92 &  $ \N = 1$ &  896 &  640 & 7\\
$(9,9)$ &  88 &  $ \N = 2$ &  1024 &  16 & 31\\
$(9,9)$ &  84 &  $ \N = 2$ &  1024 &  32 & 33\\
$(9,9)$ &  80 &  $ \N = 2$ &  1472 &  32 & 35\\
$(9,9)$ &  76 &  $ \N = 2$ &  1184 &  64 & 25\\
$(9,9)$ &  72 &  $ \N = 2$ &  1472 &  608 & 11\\
$(9,9)$ &  0 &  $ \N = 4$ &  4864 &  8 & 43\\
\hline
$(7,7)$ &  184 &  $ \N = 1$ &  704 &  64 & 10\\
$(7,7)$ &  174 &  $ \N = 1$ &  608 &  64 & 13\\
$(7,7)$ &  168 &  $ \N = 1$ &  352 &  64 & 9\\
$(7,7)$ &  166 &  $ \N = 1$ &  832 &  304 & 13\\
$(7,7)$ &  156 &  $ \N = 1$ &  736 &  64 & 18\\
$(7,7)$ &  153 &  $ \N = 1$ &  832 &  124 & 28\\
$(7,7)$ &  150 &  $ \N = 1$ &  832 &  224 & 20\\
$(7,7)$ &  144 &  $ \N = 1$ &  832 &  64 & 19\\
$(7,7)$ &  140 &  $ \N = 1$ &  800 &  296 & 15\\
$(7,7)$ &  134 &  $ \N = 1$ &  832 &  128 & 26\\
$(7,7)$ &  133 &  $ \N = 1$ &  608 &  62 & 16\\
$(7,7)$ &  132 &  $ \N = 1$ &  1184 &  688 & 15\\
$(7,7)$ &  130 &  $ \N = 1$ &  608 &  8 & 25\\
$(7,7)$ &  122 &  $ \N = 1$ &  1024 &  128 & 22\\
$(7,7)$ &  119 &  $ \N = 1$ &  1184 &  304 & 27\\
$(7,7)$ &  118 &  $ \N = 1$ &  800 &  224 & 14\\
$(7,7)$ &  116 &  $ \N = 1$ &  1184 &  592 & 18\\
$(7,7)$ &  112 &  $ \N = 1$ &  800 &  32 & 28\\
$(7,7)$ &  110 &  $ \N = 1$ &  992 &  196 & 21\\
$(7,7)$ &  109 &  $ \N = 1$ &  928 &  124 & 27\\
$(7,7)$ &  108 &  $ \N = 1$ &  784 &  224 & 9\\
$(7,7)$ &  106 &  $ \N = 1$ &  832 &  32 & 35\\
$(7,7)$ &  104 &  $ \N = 1$ &  1024 &  784 & 8\\
$(7,7)$ &  103 &  $ \N = 1$ &  832 &  62 & 38\\
$(7,7)$ &  102 &  $ \N = 1$ &  800 &  128 & 9\\
$(7,7)$ &  100 &  $ \N = 1$ &  1184 &  32 & 42\\
$(7,7)$ &  97 &  $ \N = 1$ &  704 &  62 & 18\\
$(7,7)$ &  94 &  $ \N = 1$ &  896 &  128 & 28\\
$(7,7)$ &  91 &  $ \N = 1$ &  1024 &  304 & 21\\
$(7,7)$ &  90 &  $ \N = 1$ &  1024 &  160 & 15\\
$(7,7)$ &  88 &  $ \N = 1$ &  1024 &  608 & 13\\
$(7,7)$ &  87 &  $ \N = 1$ &  1184 &  248 & 21\\
$(7,7)$ &  84 &  $ \N = 1$ &  1184 &  592 & 17\\
$(7,7)$ &  82 &  $ \N = 1$ &  992 &  160 & 15\\
$(7,7)$ &  78 &  $ \N = 1$ &  976 &  152 & 21\\
$(7,7)$ &  76 &  $ \N = 1$ &  992 &  800 & 4\\
$(7,7)$ &  72 &  $ \N = 1$ &  1024 &  32 & 32\\
$(7,7)$ &  68 &  $ \N = 1$ &  1184 &  608 & 9\\
$(7,7)$ &  60 &  $ \N = 1$ &  992 &  704 & 5\\
$(7,7)$ &  56 &  $ \N = 1$ &  976 &  800 & 3\\
\hline
$(5,5)$ &  160 &  $ \N = 2$ &  320 &  16 & 8\\
$(5,5)$ &  152 &  $ \N = 1$ &  80 &  64 & 2\\
$(5,5)$ &  148 &  $ \N = 2$ &  704 &  64 & 10\\
$(5,5)$ &  144 &  $ \N = 2$ &  896 &  16 & 18\\
$(5,5)$ &  140 &  $ \N = 2$ &  1280 &  32 & 28\\
$(5,5)$ &  136 &  $ \N = 2$ &  1600 &  128 & 20\\
$(5,5)$ &  112 &  $ \N = 1$ &  784 &  592 & 7\\
$(5,5)$ &  99 &  $ \N = 1$ &  784 &  272 & 14\\
$(5,5)$ &  96 &  $ \N = 2$ &  896 &  32 & 13\\
$(5,5)$ &  96 &  $ \N = 1$ &  784 &  592 & 6\\
$(5,5)$ &  90 &  $ \N = 1$ &  736 &  148 & 17\\
$(5,5)$ &  86 &  $ \N = 1$ &  784 &  112 & 15\\
$(5,5)$ &  84 &  $ \N = 2$ &  1024 &  64 & 22\\
$(5,5)$ &  84 &  $ \N = 1$ &  736 &  608 & 5\\
$(5,5)$ &  83 &  $ \N = 1$ &  784 &  272 & 12\\
$(5,5)$ &  80 &  $ \N = 2$ &  1472 &  8 & 25\\
$(5,5)$ &  80 &  $ \N = 1$ &  784 &  592 & 5\\
$(5,5)$ &  77 &  $ \N = 1$ &  704 &  62 & 18\\
$(5,5)$ &  76 &  $ \N = 2$ &  1792 &  128 & 30\\
$(5,5)$ &  74 &  $ \N = 1$ &  704 &  148 & 13\\
$(5,5)$ &  72 &  $ \N = 2$ &  2624 &  8 & 33\\
$(5,5)$ &  71 &  $ \N = 1$ &  704 &  296 & 9\\
$(5,5)$ &  68 &  $ \N = 1$ &  736 &  32 & 19\\
$(5,5)$ &  64 &  $ \N = 2$ &  608 &  64 & 11\\
$(5,5)$ &  62 &  $ \N = 1$ &  736 &  152 & 11\\
$(5,5)$ &  56 &  $ \N = 2$ &  704 &  16 & 22\\
$(5,5)$ &  56 &  $ \N = 1$ &  736 &  704 & 2\\
$(5,5)$ &  52 &  $ \N = 2$ &  1088 &  40 & 29\\
$(5,5)$ &  48 &  $ \N = 2$ &  1088 &  32 & 28\\
$(5,5)$ &  44 &  $ \N = 2$ &  1600 &  64 & 26\\
$(5,5)$ &  40 &  $ \N = 2$ &  1600 &  8 & 31\\
$(5,5)$ &  36 &  $ \N = 2$ &  1024 &  32 & 31\\
$(5,5)$ &  32 &  $ \N = 2$ &  832 &  32 & 9\\
$(5,5)$ &  28 &  $ \N = 2$ &  1024 &  32 & 31\\
$(5,5)$ &  24 &  $ \N = 2$ &  1408 &  32 & 28\\
$(5,5)$ &  20 &  $ \N = 2$ &  1664 &  128 & 26\\
$(5,5)$ &  16 &  $ \N = 2$ &  1984 &  32 & 23\\
$(5,5)$ &  12 &  $ \N = 2$ &  1792 &  128 & 26\\
$(5,5)$ &  8 &  $ \N = 2$ &  2624 &  608 & 14\\
\hline
$(3,3)$ &  210 &  $ \N = 1$ &  320 &  32 & 7\\
$(3,3)$ &  176 &  $ \N = 1$ &  704 &  128 & 11\\
$(3,3)$ &  148 &  $ \N = 1$ &  704 &  64 & 9\\
$(3,3)$ &  144 &  $ \N = 1$ &  608 &  64 & 11\\
$(3,3)$ &  142 &  $ \N = 1$ &  992 &  224 & 20\\
$(3,3)$ &  126 &  $ \N = 1$ &  608 &  224 & 7\\
$(3,3)$ &  114 &  $ \N = 1$ &  928 &  224 & 15\\
$(3,3)$ &  113 &  $ \N = 1$ &  640 &  124 & 14\\
$(3,3)$ &  110 &  $ \N = 1$ &  992 &  256 & 15\\
$(3,3)$ &  108 &  $ \N = 1$ &  1504 &  544 & 18\\
$(3,3)$ &  104 &  $ \N = 1$ &  608 &  32 & 16\\
$(3,3)$ &  98 &  $ \N = 1$ &  704 &  16 & 13\\
$(3,3)$ &  92 &  $ \N = 1$ &  992 &  544 & 13\\
$(3,3)$ &  86 &  $ \N = 1$ &  896 &  256 & 8\\
$(3,3)$ &  82 &  $ \N = 1$ &  928 &  128 & 10\\
$(3,3)$ &  80 &  $ \N = 1$ &  1184 &  592 & 11\\
$(3,3)$ &  79 &  $ \N = 1$ &  992 &  248 & 19\\
$(3,3)$ &  78 &  $ \N = 1$ &  992 &  224 & 9\\
$(3,3)$ &  76 &  $ \N = 1$ &  1504 &  544 & 17\\
$(3,3)$ &  70 &  $ \N = 1$ &  800 &  148 & 20\\
$(3,3)$ &  69 &  $ \N = 1$ &  736 &  124 & 12\\
$(3,3)$ &  64 &  $ \N = 1$ &  832 &  64 & 23\\
$(3,3)$ &  63 &  $ \N = 1$ &  608 &  62 & 17\\
$(3,3)$ &  60 &  $ \N = 1$ &  992 &  32 & 24\\
$(3,3)$ &  52 &  $ \N = 1$ &  1120 &  608 & 8\\
$(3,3)$ &  51 &  $ \N = 1$ &  832 &  248 & 14\\
$(3,3)$ &  48 &  $ \N = 1$ &  1184 &  592 & 10\\
$(3,3)$ &  47 &  $ \N = 1$ &  992 &  304 & 12\\
$(3,3)$ &  44 &  $ \N = 1$ &  1504 &  704 & 7\\
$(3,3)$ &  42 &  $ \N = 1$ &  800 &  128 & 12\\
$(3,3)$ &  38 &  $ \N = 1$ &  784 &  112 & 13\\
$(3,3)$ &  36 &  $ \N = 1$ &  832 &  64 & 16\\
$(3,3)$ &  32 &  $ \N = 1$ &  800 &  32 & 15\\
$(3,3)$ &  24 &  $ \N = 1$ &  1088 &  704 & 3\\
$(3,3)$ &  20 &  $ \N = 1$ &  1120 &  800 & 3\\
$(3,3)$ &  16 &  $ \N = 1$ &  1184 &  608 & 5\\
$(3,3)$ &  12 &  $ \N = 1$ &  1504 &  992 & 3\\
\hline
$(1,1)$ &  144 &  $ \N = 2$ &  64 &  16 & 3\\
\hline
\hline
\end{longtable}
\end{center}

\end{document}